\documentclass[a4paper,twoside,11pt]{article} 

\usepackage[english]{babel} %
\usepackage{lmodern}
\usepackage[T1]{fontenc}
\usepackage[latin9]{inputenc}	

\newlength{\defaultparindent}
\usepackage{latexsym}
\usepackage{amsmath}
\usepackage{amsfonts}
\usepackage{amsthm}
\usepackage{bbold}
\usepackage{multirow}
\usepackage{hyperref}
\hypersetup{colorlinks=true,citecolor=green,linkcolor= blue}
\usepackage{todonotes} 
\usepackage{soul} 
\usepackage{mathtools} 
\usepackage{rotating} 
\usepackage{arydshln} 

\usepackage[arXiv]{optional} 
\def\mynote{\todo} 

\opt{AACA}{
\def\cal{\mathcal}
}

\opt{x,AACA}{
\input{mbh_defs_TeX}
}

\opt{arXiv,std,JMP,JOPA}{

\newtheorem{MS_Proposition}{Proposition}

\def\myconjugate#1{\overline{#1}} 

\def\eg{e.g.\ }
\def\myisom{\cong} 
\newcommand{\R}{\ensuremath{\mathbb{R}}} 
\newcommand{\C}{\ensuremath{\mathbb{C}}} 
\newcommand{\F}{\ensuremath{\mathbb{F}}} 
\newcommand{\Z}{\ensuremath{\mathbb{Z}}} 
\newcommand{\HH}{\ensuremath{\mathbb{H}}} 
\newcommand{\Identity}{\ensuremath{\mathbb{1}}} 
\def\my_span#1{\mbox{Span}\left(#1\right)} 

\def\bino#1#2{\left( \! \begin{array}{c} #1 \\ #2 \end{array} \! \right)}
\def\dotinformula{\;\; \mathrm{.}} 

\def\End{\ensuremath{\mbox{End}}}

\newcommand{\anticomm}[2]{\ensuremath{\left\{ #1, #2 \right\}}} 
\newcommand{\myClf}[3]{\ensuremath{{{\cal C}\ell}_{#3} {\left( #1, #2 \right)}}}	
\newcommand{\myClC}[3]{\ensuremath{{{\cal C}\ell}_{\C} {\left( #1 \right)}}}	
\newcommand{\myClg}[3]{\ensuremath{{{\cal C}\ell} {\left( #3 \right)}}}	

}

\opt{JMP}{
}

\def\h_eigen{\eta}
\def\g_eigen{\theta}
\def\mygen{e} 
\def\mydual{{*}} 
\def\mygenC{f} 

\newcommand{\Mod}[1]{\ (\textup{mod}\ #1)}
\def\mymod{\Mod} 

\begin{document}

\opt{x,std,arXiv,JMP,JOPA}{
\title{{\bf
On complex representations of Clifford algebra 
} 
	}

\author{\\
	\bf{Marco Budinich}%
%
%
\\
	University of Trieste and INFN, Trieste, Italy\\
	\texttt{mbh@ts.infn.it}\\
%
%
%
	}
\date{ \today }
\maketitle
}

\opt{AACA}{
\title{On space signatures}

\author{Marco Budinich}
\address{Dipartimento di Fisica\\
	Università di Trieste \& INFN\\
	Via Valerio 2, I - 34127 Trieste, Italy}
\email{mbh@ts.infn.it}
}

\begin{abstract}
We show that complex representations of Clifford algebra can always be reduced either to a real or to a quaternionic algebra depending on signature of complex space thus showing that complex spinors are unavoidably either real Majorana spinors or quaternionic spinors. We use this result to support $(1,3)$ signature for Minkowski space.
\end{abstract}


\opt{AACA}{
\keywords{Clifford algebra; binary integers; spinors; periodicity.}
\maketitle
}

\section{Introduction}
\label{Introduction}
\opt{margin_notes}{\mynote{mbh.note: for paper material see log pp. 703--712 \& relative references}}%
In 1913 {\'{E}}lie Cartan introduced spinors \cite{Cartan_1913, Cartan_1937} and, after more than a century, their offsprings continue to blossom.
Spinors were later thoroughly investigated by Claude Chevalley \cite{Chevalley_1954} in the mathematical frame of Clifford algebras and identified as elements of minimal left ideals of the algebra.

For a real vector space $V \myisom \R^{k, l}$ the properties of its Clifford algebra $\myClg{}{}{V}$ are fully determined by its signature $(k, l)$ and it is customary to define a bijective transformation in this plane: $n := k + l$ and $\nu := k - l$.

\opt{margin_notes}{\mynote{mbh.note: in Lounesto book (history at the end) it's written that is Cartan that first discovered the periodicities that are frequently ascribed to Bott!}}%
The reason behind this transformation is that $\nu \mymod{8}$ determines uniquely both the underlying division algebra and to which of the 8 classes of the Brauer--Wall group of $\R$ the Clifford algebra belongs \cite[p.~89]{BudinichP_1988e}, while the symmetry properties of the invariant bilinear forms of spinor space depend on $n \mymod{8}$. $n$ and $\nu$ together give rise to the ``spinorial chessboard'' that appears in the title of the book \cite[p.~109]{BudinichP_1988e}. In binary form $\nu, n \mymod{8}$ have just 3 bits and in \cite{Budinich_2016b} we have shown that it is wiser to consider these bits since they correspond to precise properties of $\myClg{}{}{V}$ tightly intertwined with the main involutions of the algebra.
%
%
%

In this paper we begin generalizing \cite{Budinich_2016b} to the case of a complex space $V$ and its $\myClg{}{}{V}$ and with this new result we show that the complex $\myClg{}{}{V}$ is always reducible either to a real algebra with Majorana spinors or to a quaternionic algebra and thus that complex numbers are not really needed.

%
%
%
%
%
%
%

More in detail in section~\ref{Real_Clifford_algebra} we start by some remarks on binary numbers and then introduce the main argument resuming the results of \cite{Budinich_2016b} that show how the main involutions of the real $\myClg{}{}{V}$ fully determine the relations between the bits of $\nu$ and $n$ and get rid of all periodicities of $\myClg{}{}{V}$. In following section~\ref{Generalization_to_C} we prove that this result neatly generalize to the case of complex spaces and the relative Clifford algebras simply extending the set of main involutions with the involution related to complex conjugation; for the sake of completeness in section~\ref{Relations_with_C_representations} these results are related to the similar ones obtained for complex representations of Clifford algebras \cite{Trautman_1998}; this part is not needed for what follows. In last section~\ref{CCaiar} we apply these results to investigate the ubiquitous case of complex representations of Clifford algebras and we show that it is always possible to see a complex Clifford algebra as a tensor product complexification and thus that the complex Clifford algebra can always be reduced to one of its subalgebras. This is well known for Majorana spinors but we show it is possible to do it also in the other possible case when Clifford algebra and spinors become quaternionic and thus in both cases we can eliminate complex numbers. This in turn allows to argue in favour of signature $(1,3)$ for Minkowski space.

For the convenience of the reader we tried to make this paper as elementary and self-contained as possible.

\section{The case of real Clifford algebra}
\label{Real_Clifford_algebra}
\opt{margin_notes}{\mynote{mbh.ref: for this part see pp. 650 ff.}}%
We resume some simple facts on the binary representation of an integer $n$:
\begin{equation}
\label{binary_expansion}
n = \left\{
\begin{array}{l l}
\sum_{i = 0}^\infty n_i 2^i & n_i \in \{ 0, 1 \} \\
\sum_{i = 0}^\infty \frac{1 - \prescript{}{i}n}{2} 2^i & \prescript{}{i}n \in \{ 1, -1 \}, \quad \prescript{}{i}n = (-1)^{n_i}
\end{array} \right.
\end{equation}
and we will switch between the two forms $n_i$ and $\prescript{}{i}n$ of the binary expansion as and when it suits to us.


Representing an integer $n$ with a finite number $k$ of bits one is de facto implementing modular arithmetic, namely $n \mymod{2^k}$. In this case the customary ``2-complement'' representation for negative numbers is given by $-n = 2^k - n$, that satisfies $n + (-n) \equiv 0 \mymod{2^k}$. By Clifford algebra periodicities many algebra properties depend on integers modulo $8$, thus meaning that just the three least significant bits of integer $n$ are relevant so that we will frequently use $n' := n \mymod{8}$ and it is simple to verify that, if $n' \ne 0$ then $- n \mymod{8} = 8 - n'$.

A remarkable result descending from Lucas' theorem (see \eg \cite{Granville_1997}) is
\begin{equation}
\label{Lucas_theorem}
\prescript{}{i}n = (-1)^{\bino{n}{2^i}} \qquad (n \; \mbox{integer})
\end{equation}
that shows that the formulas $(-1)^{\frac{n (n - 1)}{2}}$ and $(-1)^n$, that frequently occur in Clifford algebras, are just the two least significant bits $\prescript{}{1}n$ and $\prescript{}{0}n$ of the binary representation of $n$.

%
%
\begin{table}
\centering
\begin{tabular}{c | c | c c c c c c c c |}
\multicolumn{2}{c}{} & \multicolumn{8}{c}{$\nu$} \\
\cline{3-10}
\multicolumn{1}{c}{} & & $0$ & $1$ & $2$ & $3$ & $4$& $5$& $6$& $7$ \\
\cline{2-10}
\multirow{8}{*}{$n$}
%
%
%
%
& $0$ & $\R$ & & & & & & & \\
& $1$ & & $2 \R$ & & & & & & \\
& $2$ & $\R(2)$ & & $\R(2)$ & & & & & \\
& $3$ & & $2 \R(2)$ & & $\C(2)$ & & & & \\
& $4$ & $\R(4)$ & & $\R(4)$ & & $\HH(2)$ & & & \\
& $5$ & & $2 \R(4)$ & & $\C(4)$ & & $2 \HH(2)$ & & \\
& $6$ & $\R(8)$ & & $\R(8)$ & & $\HH(4)$ & & $\HH(4)$ & \\
& $7$ & & $2 \R(8)$ & & $\C(8)$ & & $2 \HH(4)$ & & $\C(8)$ \\
\cline{2-10}
\end{tabular}
\caption{The real, universal, Clifford algebras of $V = \R^{k,l}$ depending on $n = k + l$ and $\nu = k - l$. $\R(2)$ is the matrix algebra of real $2 \times 2$ matrices and $2 \R$ stands for $\R \oplus \R$; for negative values of $\nu$ the division algebra can be derived remembering that, for modulo $8$ values, $- \nu = 8 - \nu$.}
\label{n_nu_table_reduced}
\end{table}%
In the literature \cite{BudinichP_1988e, Porteous_1981} it is customary to arrange the periodicity properties of Clifford algebras in a table of $n$ vs $\nu$, like in table~\ref{n_nu_table_reduced}, or $- \nu$ like in the the spinorial clock of \cite[p.~122]{BudinichP_1988e}%
%
%
%
%
%
.
\opt{margin_notes}{\mynote{mbh.ref: About the Babel of ...tables see log. p. 640'; the spinorial clock appears also in \cite[p.~7]{Trautman_1998} here a long footnote is commented}}%

In \cite{Budinich_2016b} we have shown that it is instructive to arrange the various kinds of Clifford algebras (more precisely: the division algebras of the matrix algebra of Wedderburn-Artin theorem) as depending on the three least significant bits of $\nu$ and we resume their meaning with
\begin{equation}
\label{nu_bits_def}
\begin{array}{l l l}
\prescript{}{2}\nu & = & \left\{
\begin{array}{l l}
1 & \quad \R \\
-1 & \quad \HH
\end{array} \right. \\
\prescript{}{1}\nu & = & \omega^2 \\
\prescript{}{0}\nu & = & \left\{
\begin{array}{l l}
1 & \quad \nu, n \; \mbox{even} \qquad \mbox{(central, simple algebra)} \\
-1 & \quad \nu, n \; \mbox{odd} \qquad \; \mbox{(}\omega \in \; \mbox{center)}
\end{array} \right.
\end{array}
\end{equation}
where $\omega = \mygen_1 \mygen_2 \cdots \mygen_n$ is the product of the generators, namely the volume element and for Clifford algebras over a real vector space $V = \R^{k, l}$ it is a standard exercise to calculate $\omega^2$ and using (\ref{Lucas_theorem})
$$
\omega^2 = (-1)^{\frac{(k - l)(k - l - 1)}{2}} = (-1)^{\bino{\nu}{2}} = \prescript{}{1}\nu \dotinformula
$$

\begin{figure}
\centering
\includegraphics[scale=0.4]{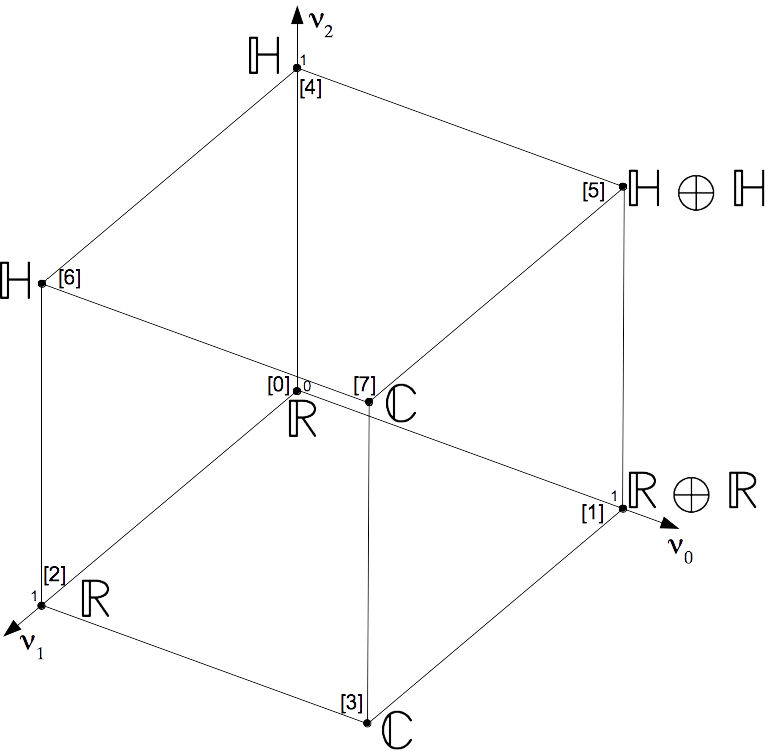}
%
%
%
\caption{The $5$ different division algebras of the real Clifford algebras of $V = \R^{k,l}$ as depending on the three least significant bits $\nu_2$, $\nu_1$ and $\nu_0$ of the binary representation of $\nu = k - l$, the number in square brackets near each cube vertex.}
\label{nu_drawing}
\end{figure}

Exploiting this result it is evocative to replace table~\ref{n_nu_table_reduced} with a 3-dimensional cube with the 3 axes corresponding respectively to $\nu_0, \nu_1$ and $\nu_2$ and with each division algebra on its corresponding vertex as done in figure~\ref{nu_drawing}. We see that the cube face $\nu_0 = 0$ ($\nu, n$ even) contains all central simple algebras whereas the face $\nu_0 = 1$ ($\nu, n$ odd) contains all non central algebras for whom the center is $\{ \Identity, \omega \}$. The two faces $\nu_1 = 0, 1$ correspond respectively to $\omega^2 = 1, -1$ and finally the faces $\nu_2 = 0, 1$ contain respectively $\R$ or $\HH$ division algebras.

The vertical edge $\nu_0 = \nu_1 = 1$ contains the cases in which $\omega$ belongs to the center and $\omega^2 = -1$ and in these cases the center is isomorphic to $\C$. On the other edge $\nu_0 = 1, \nu_1 = 0$ there are the cases in which $\omega$ belongs to the center but $\omega^2 = 1$, namely the cases of ``double'' algebra $\R \oplus \R$ and $\HH \oplus \HH$.

The advantage of this representation is that each axis of the cube represent a unique, well defined, property of the algebra independent of all others, thus superseding traditional $\nu \mymod{4}, \nu \mymod{2}$ characterizations; this renders the properties of Clifford algebras easier to visualize.

Given $\nu$ we learn something about the Clifford algebra but this is not the entire story since different possible alternatives for $n$ remain. This further ``degree of freedom'' is responsible of other periodicities of Clifford algebra that give rise to the spinorial chessboard \cite[p.~109]{BudinichP_1988e} and surfaces in a different form producing the eight double coverings of the group O, the Dabrowski groups \cite{Dabrowski_1988}. Dabrowski defines three variables, named $a, b, c \in \{ \pm 1 \}$, that completely determines the characteristics of Clifford algebra; his work has been subsequently developed by Varlamov \cite{Varlamov_2001}.

We have shown \cite{Budinich_2016b} that this degree of freedom depends on the fundamental automorphisms of Clifford algebras and we resume the story here beginning from the fundamental automorphisms of Clifford algebras. In general in Clifford algebras there are four automorphisms corresponding to the two involutions and to the two antinvolutions induced by the orthogonal involutions $\Identity_V$ and $-\Identity_V$ of vector space $V$ \cite[Theorem~13.31]{Porteous_1981}. They are called fundamental or discrete automorphisms and under composition form a finite group, isomorphic to $\Z_2 \otimes \Z_2$ \cite{Varlamov_2001}.

In what follows we treat the simpler case of even $n$ leaving the odd case for future analysis. For $n$ even the Clifford algebra is central simple and, by Skolem--Noether theorem, all its automorphisms are inner. The inner elements $\omega$ and
\begin{equation}
\label{tau_def}
\tau := \left\{
\begin{array}{l l}
\mygen_{k+1} \mygen_{k + 2} \cdots \mygen_{k + l} & \quad \mbox{for} \; k,l \; \mbox{even} \\
\mygen_{1} \mygen_{2} \cdots \mygen_{k} & \quad \mbox{for} \; k,l \; \mbox{odd}
\end{array} \right.
\end{equation}
(here $\mygen_i^2 = 1$ for $i \le k$ and $\mygen_i^2 = -1$ for $i > k$) give \cite{Budinich_2016}:
\begin{equation}
\label{fundamental_automorphisms}
\begin{array}{l l r l l}
\omega \mygen_i \omega^{-1} & = & - \mygen_i & & \qquad \qquad \forall \: 1 \le i \le n \\
\tau \mygen_{i} \tau^{-1} & = & \mygen_{i}^{\mydual} & = & \mygen_{i}^{-1} 
\end{array}
\end{equation}
where $\mygen_{i}^{\mydual}$ is the dual of $\mygen_{i}$. Together with their composition $\omega \tau$ and identity they define the 4 fundamental involutions of real Clifford algebras (not to be mistaken with antinvolutions, \eg reversion) and form group $\Z_2 \otimes \Z_2$.

The two bits $\tau^2 = \pm 1$ and $s_1 = \pm 1$ defined by
\begin{equation}
\label{s1_def}
\omega \tau \omega^{-1}= s_1 \tau
\end{equation}
fully define the relations between the bits of $\nu$ and the corresponding bits of $n$%
\opt{margin_notes}{\mynote{mbh.ref: log p. 646}}%
: for $n$ even the following relations hold \cite{Budinich_2016b}:
\begin{equation}
\label{n_bits_def}
\begin{array}{l l l l l}
\prescript{}{2}\nu & = & \prescript{}{2}n \, \tau^2 \\
\prescript{}{1}\nu & = & \prescript{}{1}n \, s_1 & = & \omega^2 \\
\prescript{}{0}\nu & = & \prescript{}{0}n & = & 1
\end{array}
\end{equation}
and thus knowing \eg $\nu$ and the two additional bits $\tau^2$ and $s_1$ we can determine also $n \mymod{8}$ thus removing all periodicities from Clifford algebra. In Varlamov's notation \cite{Varlamov_2001} $(a, b, c) = (s_1 \omega^2 \tau^2, \tau^2, \omega^2) = (\prescript{}{1}n \prescript{}{2}n \prescript{}{2}\nu, \prescript{}{2}n \prescript{}{2}\nu, \prescript{}{1}\nu)$.

\section{Generalization to $\C$ case}
\label{Generalization_to_C}
To generalize (\ref{n_bits_def}) to $\C$ case we first need to define properly the meaning of $\nu$ in this case since it is well known that in complex quadratic space $V \myisom \C^n$ to any orthonormal base we can assign any signature multiplying base vectors by a field coefficient $\iota \in \{1, i \}$.

The Clifford algebra of any quadratic space $V$ contains ``copies'' of the field $\F$ and of $V$ by injections $\F \to \myClg{}{}{V}: x \to x \Identity$ and $V \to \myClg{}{}{V}$, the so called Clifford map. Whereas the field injection is unique, there are various possibilities for the injections of vector space $V$. In any case the Clifford map always selects a subset of the anticommuting matrices of ``its'' matrix algebra $\mygen_j$ and it is customary to identify $V$ with its image in $\myClg{}{}{V}$ and an orthonormal base of $V$ with the anticommuting matrices $\mygen_j$ (in the real case this is the only available choice). We remark that these matrices are not ``representations'' but members of the algebra itself.

In the complex case the generators $\mygen_j \in \myClg{}{}{V}$ can be redefined multiplying them by the field coefficient $\iota \in \{1, i \}$ and we name them $\mygenC_j$ to underline the difference with the real case
\begin{equation}
\label{f_j_def}
\mygenC_j := \iota_j \mygen_j
\end{equation}
the sign of any $\mygenC_j^2 = \pm 1$ can be chosen at will and thus $V$ signature.

Only after we have done these two choices: the Clifford map and the field coefficients $\iota_j$ a signature is well defined and there will be $k$ space like generators ($\mygenC_j^2 = 1$) and $l = n - k$ timelike ones ($\mygenC_j^2 = -1$) and we show that doing these choices we left a footprint in the algebra.

Once the signature is freezed all the homogeneous elements (blades), as for example the volume element that now we write $\omega = \mygenC_1 \mygenC_2 \cdots \mygenC_n$, have a ``natural'' value for their square, e.g. $\omega^2$ descending from the values of $\mygenC_j^2$. It is true that we can redefine also $\omega' = i \omega$ and $\omega'^2 = - \omega^2$ but this is only cosmetics since neither the involution automorphism induced by $\omega$ is altered nor is altered the natural sign of its square nor are altered other properties like \eg the sign $s_1$ (\ref{s1_def}) and so from now on we will always consider for $\omega^2$ and other blades what we call their \emph{natural} sign.

In complex space the set of generators $\mygenC_j$ has two natural partitions into two classes: the first depends on $\iota_j$ giving $\myconjugate{\mygenC_{j}} = \pm \mygenC_{j}$, the second on $\mygenC_j^{-1} = \pm \mygenC_j$ (we just remind that also in the complex case $\mygenC_j^{-1} = \mygenC_j^3 = \pm \mygenC_j$). These two partitions lift to two involutions and, if the dimension of the linear space $V$ is even, by Skolem--Noether theorem, there exist corresponding inner elements. For the first partition there must exists $\rho$ such that
\begin{equation}
\label{rho_def}
\rho \mygenC_{j} \rho^{-1} = \myconjugate{\mygenC_{j}} = \pm \mygenC_{j}
\end{equation}
and we remark that this is a $\C$ linear operation and thus \emph{is not} complex conjugation but \emph{coincides} with it for the generators since $\iota_j \in \{1, i\}$ and for them complex conjugation reduces to a sign; (\ref{rho_def}) defines a third fundamental involution of $\myClg{}{}{V}$ that, together with the two of the real case (\ref{fundamental_automorphisms}), give rise to a finite group with 8 elements. Supposing that there are $0 \le r \le n$ field coefficients $\iota_j = i$ and $n - r$ $\iota_j = 1$ we see that (\ref{rho_def}) is satisfied by:
\begin{equation}
\label{rho_explicit_def}
\rho := \left\{
\begin{array}{l l}
\prod_{j \in \{r\}} \mygenC_j & \quad \mbox{for} \; r \; \mbox{even} \\
\prod_{j \in \{n - r\}} \mygenC_j & \quad \mbox{for} \; r \; \mbox{odd}
\end{array} \right. 
\end{equation}
where by $j \in \{r\}$ we mean the set of generators for which $\iota_j = i$ and by $j \in \{n - r\}$ we mean the complementary set for which $\iota_j = 1$. For the second partition there must exist an algebra element $\Theta$ such that
\begin{equation}
\label{Theta_def}
\Theta \mygenC_j \Theta^{-1} = \mygenC_j^{-1} = \pm \mygenC_{j}
\end{equation}
and we easily verify that
\begin{equation}
\label{Theta_def_2}
\Theta := \left\{
\begin{array}{l l}
\tau & \quad \mbox{for} \quad \R \quad \mbox{and} \quad \mygen_j^{-1} = \mygen_{j}^{\mydual} \\
\tau \rho & \quad \mbox{for} \quad \C \quad \mbox{and} \quad \mygenC_j^{-1} = \myconjugate{\mygenC_{j}}^{\mydual} \dotinformula
\end{array} \right. 
\end{equation}
With these definitions we may finally generalize (\ref{n_bits_def}) with

\begin{MS_Proposition}
\label{Theta_prop}
In any Clifford algebra of a real or complex quadratic space $V$ of even dimension $n = 2m$ and signature $(k, l)$ let $\Theta$ be the inner element giving the inverse of the generators (\ref{Theta_def}) and let $s = \pm 1$ be defined by
$$
\omega \Theta \omega^{-1} = s \Theta
$$
then the three bits of $\nu = k - l$ of $V$ are given by
\begin{equation}
\label{general_nu_bits_derivation}
\begin{array}{l l l l l}
\prescript{}{2}\nu & = & \prescript{}{2}n \, \Theta^2 \\
\prescript{}{1}\nu & = & \prescript{}{1}n \, s & = & \omega^2 \\
\prescript{}{0}\nu & = & \prescript{}{0}n & = & 1
\end{array}
\end{equation}
\end{MS_Proposition}

\noindent which tedious proof is in the Appendix. Three remarks on this result: the first is that given the dimension $n$ of $V$ and the value of the bits $\omega^2$ (or $s$) and $\Theta^2$ one obtains $\nu$ and thus the complete signature also in complex case.

The second remark is that in passing from the real to the complex case we need to add $\rho$ to the definition of $\Theta $ (\ref{Theta_def_2}); similarly with bits $\rho^2$ and $s_2$ (\ref{s_i_defs}) we go from (\ref{n_bits_def}) to (\ref{general_nu_bits_derivation}) (further details in the proof)
\begin{subequations}\label{general_nu_bits_derivation_2}
\begin{alignat}{2}
\label{general_nu_bits_derivation_2a}
\prescript{}{2}\nu & = \prescript{}{2}\nu_\R \, \rho^2 & = \prescript{}{2}n \, \tau_\R^2 \rho^2 \\
\prescript{}{1}\nu & = \prescript{}{1}\nu_\R \, s_2 & = \prescript{}{1}n \, s_1 s_2 & = \omega^2 \\
\prescript{}{0}\nu & = \prescript{}{0}\nu_\R & = \prescript{}{0}n \qquad & = 1 
\end{alignat}
\end{subequations}
where with $\nu_\R$ and $\tau_\R$ we indicate respectively $\nu$ and $\tau$ we would get in the real case or, equivalently, from $\mygenC_j$ setting all $\iota_j = 1$ (giving $\rho = \Identity, s_2 = 1$).

The last remark is that proposition~\ref{Theta_prop} is an extension of \cite[proposition~1]{Trautman_1998} that can be derived from (\ref{general_nu_bits_derivation_2a}) in the particular case of $\prescript{}{2}\nu_\R = 1$, more on this in the following section.

\section{Relations with complex representations of \myClg{}{}{V}}
\label{Relations_with_C_representations}
\opt{margin_notes}{\mynote{mbh.ref: for $\tau B$ puzzle see log. pp. 651-2, 652.8 and 706-7}}%
In this section, not strictly necessary for what follows, we examine in detail the relations intercurring between inner involutions
\begin{equation}
\label{inner_involutions}
\begin{array}{l l l}
\mygenC_j^{\mydual} & = & \tau \mygenC_j \tau^{-1} \\
\myconjugate{\mygenC}_j & = & \rho \mygenC_j \rho^{-1}
\end{array}
\end{equation}
and the more familiar complex representations of Clifford algebras \cite{BudinichP_1988e, Trautman_1998}.

Given an even dimensional real or complex quadratic space $V$, $n = 2m$, one can build a representation of its Clifford algebra that is an algebra map
\begin{equation}
\label{representations}
\gamma: \myClg{}{}{V} \to \End_\C S
\end{equation}
in the complex linear space $S \myisom \C^{2^m}$, the space of spinors. One then defines the maps $B: S \to S^{\mydual}$ and $C: S \to \myconjugate{S}$ intertwining the equivalent complex representations $\gamma$, $\gamma^{\mydual}$ and $\myconjugate{\gamma}$ and we will concentrate on the relations between these maps and inner involutions (\ref{inner_involutions}). Maps $B$ and $C$ give the ``representation version'' of the involutions (\ref{inner_involutions})
\begin{equation}
\label{representation_involutions}
\begin{array}{l l l}
\gamma^{\mydual}(\mygenC_{j}) & = & B \gamma(\mygenC_{j}) B^{-1} \\
\myconjugate{\gamma}(\mygenC_{j}) & = & C \gamma(\mygenC_{j}) C^{-1}
\end{array}
\end{equation}
and sport the properties \cite[proposition~1]{Trautman_1998}
\begin{equation}
\label{Trautman_properties_1}
\begin{array}{l l l}
B^{\mydual} & = & (-1)^{\frac{m (m - 1)}{2}} B \\
\myconjugate{C} C & = & (-1)^{\frac{(l - k) (l - k + 2)}{8}}
\end{array}
\end{equation}
and we observe that $(-1)^{\frac{m (m - 1)}{2}} = \prescript{}{1}m = \prescript{}{2}n$ since $n = 2m$. Moreover $(-1)^{\frac{(l - k) (l - k + 2)}{8}} = (-1)^{\frac{1}{2} \frac{\nu}{2} (\frac{\nu}{2} - 1)} = \prescript{}{1}(\frac{\nu}{2}) = \prescript{}{2}\nu$ since $\nu$ is even and thus we rewrite (\ref{Trautman_properties_1}) as
\begin{equation}
\label{Trautman_properties_2}
\begin{array}{l l l}
B^{\mydual} & = & \prescript{}{2}n B \\
\myconjugate{C} C & = & \prescript{}{2}\nu \dotinformula
\end{array}
\end{equation}
Comparing (\ref{inner_involutions}) with (\ref{representation_involutions}) we realize that maps $B$ and $C$ are simply the representations of the corresponding inner elements, namely $B = \gamma(\tau)$ and $C = \gamma(\rho)$, since their actions on the generators is the same.

By (\ref{tau_def}) and (\ref{f_j_def}), if $\tau$ is made by $x \ge 0$ generators, $B = \gamma(\tau) = \gamma(\mygenC_{j_1}) \cdots \gamma(\mygenC_{j_x})$ then
$$
B^{\mydual} = \gamma^{\mydual}(\mygenC_{j_x}) \cdots \gamma^{\mydual}(\mygenC_{j_1}) = \gamma(\mygenC_{j_x}) \cdots \gamma(\mygenC_{j_1}) 
$$
since in both cases of (\ref{tau_def}) the transformation from dual to plain form leaves a factor $1$. Thus reordering the generators we get $B$ and so
$$
B^{\mydual} = (-1)^\frac{x (x-1)}{2} B = \prescript{}{1}x B = \tau_\R^2 B
$$
since $\prescript{}{1}x$ is either $\prescript{}{1}l$ or $\prescript{}{1}k$ (\ref{tau_def}) and is precisely the value of $\tau_\R^2$ \cite[(17)]{Budinich_2016b}. For the second relation $\myconjugate{C} C = \myconjugate{\gamma(\rho)} \gamma(\rho) = \myconjugate{\gamma}(\rho) \gamma(\rho)$ by the definition of the representation $\myconjugate{\gamma}$. By (\ref{representation_involutions}) $\myconjugate{\gamma}(\rho) \gamma(\rho) = \gamma(\rho) \gamma(\rho) = \gamma(\rho^2) = \rho^2$
\opt{margin_notes}{\mynote{mbh.ref: proposition with old, longer, proof in section~\ref{Unused_parts}}}%
and so from (\ref{Trautman_properties_2}) we get
\begin{equation}
\label{Trautman_properties_3}
\begin{array}{l l l}
\tau_\R^2 & = & \prescript{}{2}n \\
\rho^2 & = & \prescript{}{2}\nu
\end{array}
\end{equation}
that allow to compare (\ref{Trautman_properties_1}) with the results of proposition~\ref{Theta_prop}. The first thing to observe is that the two relations of (\ref{Trautman_properties_3}) are equivalent since substituting each of them in (\ref{general_nu_bits_derivation_2a}) we obtain the other.

Comparing $\rho^2 = \prescript{}{2}\nu$ with (\ref{general_nu_bits_derivation_2a}) we see that they agree only iff $\prescript{}{2}\nu_\R = 1$. This apparent mismatch arises because $\gamma$ (\ref{representations}) is a representation acting on complex space $S$ whereas (\ref{general_nu_bits_derivation_2}) are \emph{always} valid, also when the algebra is quaternionic and $\prescript{}{2}\nu_\R = -1$ (\ref{nu_bits_def}).

Let us explain this with a simple example: let $V = \R^{0,2}$, then $\myClg{}{}{\R^{0,2}} \myisom \HH$ and generators $\mygen_1, \mygen_2$ can be taken to be \eg $i, j \in \HH$ and the inner element $\tau = k \in \HH$ gives
$$
\tau \mygen_i \tau^{-1} = \mygen_i^\mydual = - \mygen_i \qquad \mbox{for} \; i = 1,2
$$
moreover $\tau^2 \equiv \tau_\R^2 = -1$ that satisfies (\ref{general_nu_bits_derivation_2a}) since here $\rho^2 = 1$ and so $\prescript{}{2}\nu_\R = \prescript{}{2}n \tau_\R^2 = -1$. Going to the complex representation of $\myClg{}{}{\R^{0,2}}$ in $S = \C^2$ we can choose for the two generators \eg $\mygenC_1 = i \left(\begin{array}{r r} 0 & 1 \\ 1 & 0 \end{array}\right), \mygenC_2 = \left(\begin{array}{r r} 0 & -1 \\ 1 & 0 \end{array}\right) \in \C(2)$ and the inner elements are respectively $\tau = \mygenC_1$ and $\rho = \mygenC_2$ that give
$$
\begin{array}{l l l l r l l l l l r}
\tau \mygenC_1 \tau^{-1} & = & \mygenC_1^\mydual & = & \mygenC_1 & \qquad \qquad & \rho \mygenC_1 \rho^{-1} & = & \myconjugate{\mygenC}_1 & = & - \mygenC_1 \\
\tau \mygenC_2 \tau^{-1} & = & \mygenC_2^\mydual & = & - \mygenC_2 & \qquad \qquad & \rho \mygenC_2 \rho^{-1} & = & \myconjugate{\mygenC}_2 & = & \mygenC_2 
\end{array}
$$
\opt{margin_notes}{\mynote{mbh.note: here to check with (\ref{general_nu_bits_derivation_2a}) one needs $\tau_\R^2$ but we have only $\tau^2 = -1$ and would need $s_3$ that is defined in the proof and (\ref{s_i_defs})...}}%
and $\rho^2 = -1$ that satisfy (\ref{general_nu_bits_derivation_2a}) but now $\prescript{}{2}\nu_\R = 1$ since in the complex representation $\C(2)$ the dual structure $\mygen_i^\mydual = - \mygen_i$ of $\HH$ is vanished.

So going to complex representations of quaternionic algebras some structure is lost and $B^\mydual$ (\ref{Trautman_properties_1}) is correct. This happens because the dual (loosely: transpose) structure of complex and real matrices is \emph{identical} whereas the dual structure of quaternions is different and get lost when going to complex representations and this gives rise to the apparent mismatch between (\ref{general_nu_bits_derivation_2}) and (\ref{Trautman_properties_3}).

We conclude observing that the Hermiticity of the map $\myconjugate{B} C$ \cite[proposition~1]{Trautman_1998} is straightforward to prove in our settings:
$$
(\myconjugate{\tau} \rho)^\dagger = \tau \rho (\myconjugate{\tau} \rho) \rho^{-1} \tau^{-1} = \tau \rho (\rho \tau \rho^{-1} \rho) \rho^{-1} \tau^{-1} = \tau^2 \rho^2 \rho^{-1} \tau^{-1} = \rho \tau = \myconjugate{\tau} \rho \dotinformula
$$

\section{Complex Clifford algebra is always reducible}
\label{CCaiar}
It is well known that for a real quadratic space $V$ its signature determines the Brauer--Wall class of its Clifford algebra \myClg{}{}{V}: these relations appear in figure~\ref{nu_drawing} and in (\ref{nu_bits_def}) and (\ref{n_bits_def}). On the other hand if $V$ is complex of dimension $2m$ then $\myClg{}{}{V} \myisom \C(2^m)$ independently of the signature of $V$, that can be chosen at will. This is usually interpreted saying that in complex space the signature does not affect the Clifford algebra of $V$.

In section~\ref{Generalization_to_C} we have defined a signature for $V$ as a two step process: the choice of the Clifford map and  the choice of the coefficients $\iota_j$ assigned to generators (\ref{f_j_def}). We show that also in complex space the signature of $V$ plays a role since it \emph{selects} a subalgebra of $\C(2^m)$ to which we can always ``reduce''. This is well known for the case of reduction to real representations and to Majorana spinors. We show that this is only part of the story and that the Clifford algebra can always be ``reduced'', albeit in a generalized sense, either to a real or to a quaternionic representation and that this depends on the signature $\nu$ of complex space $V$, like in the real case.

Complex representations of Clifford algebra are of paramount importance in physics being essential for electromagnetism \cite{Trautman_1998}. Complex representation often can be seen either as an algebra map $\gamma$ (\ref{representations}) or as a complexification $\C \otimes \myClg{}{}{\R^{k,l}}$; our preference goes to the latter.
\opt{margin_notes}{\mynote{mbh.ref: see log p. 652.4}}%

We review the familiar reduction of complex spinors to real Majorana spinors to introduce a generalized definition of reduction: given $V$ with signature $\nu \mymod{8} \in \{0, 2\}$ and looking at the complex representation (\ref{representations}) the complex space of spinors $S = \C^{2^m} \myisom \R^{2^m} \oplus \R^{2^m}$ can be reduced to one of its real subspaces $\R^{2^m}$ since the representation $\gamma$ leaves invariant both of them \cite{BudinichP_1988e} and thus is reducible in the usual meaning: reduction to an invariant subspace of $S$.
\opt{margin_notes}{\mynote{mbh.ref: the two real representations are real equivalent \cite[pp.~50 and 91]{BudinichP_1988e}}}%
On the other hand we have the fully general relation
\begin{equation}
\label{C_R_split}
\C(2^m) \myisom \R(2^m) \oplus \R(2^m) \myisom \C \otimes \R(2^m)
\end{equation}
\opt{margin_notes}{\mynote{mbh.note: or $\C(2^m) \myisom \R(2^m) + i \R(2^m)$}}%
that hints at the two equivalent paths to complexify a real linear space \cite{Conrad_2008}. The representation reduced to Majorana spinors clearly refers to the first of these two isomorphisms. But we can derive it also from the second isomorphism: the real subspaces of $S$ are invariant for representation $\gamma$ that results thus isomorphic to $\R(2^m)$; in this fashion we see reduction as a reduction of the algebra $\C(2^m)$ to its subalgebra $\R(2^m)$ appearing in the second isomorphism and in this case the two definitions are equivalent.
\opt{margin_notes}{\mynote{mbh.note: to be proved ? How does Clifford algebra universality enter the business ?}}%

The image of $\myClg{}{}{V}$ by $\gamma$ (\ref{representations}) is always a subalgebra of $\End_\C S$, possibly over a different field as in the case of Majorana spinors. If this subalgebra has real dimension lower than that of $\End_\C S$ we define it a reducible representation. In the real case also the minimal left ideal $\R^{2^m}$ is a subspace of the original $\C^{2^m}$ and of halved dimension but we will show that with this generalized definition of reducibility this is not always true.

To apply the same procedure also in the quaternionic case we need
$$
\C(2^m) \myisom \HH(2^{m-1}) \oplus \HH(2^{m-1})
$$
\opt{margin_notes}{\mynote{mbh.ref: full proposition with proof in section~\ref{Unused_parts}}}%
simple to prove by induction on $m$ observing that for $m = 1$ the generic element of $\C(2)$ has the form $\left(\begin{array}{r r} a & b \\ c & d \end{array}\right) \myisom \left(\begin{array}{r r} a & 0 \\ c & 0 \end{array}\right) \oplus \left(\begin{array}{r r} 0 & b \\ 0 & d \end{array}\right)$ each term being isomorphic to $\C^2$ and thus to $\HH$ via the standard isomorphism \cite{Porteous_1995}
\begin{equation}
\label{C_to_H_isomorphism}
\C^n \times \C^n \to \HH^n; (z_1,z_2) \to z_1 + j z_2 \dotinformula
\end{equation}
With this result and with the standard one $\C(2^m) \myisom \C \otimes \HH(2^{m-1})$ \cite{Porteous_1995} we can write a relation similar to (\ref{C_R_split})
\begin{equation}
\label{C_H_split}
\C(2^m) \myisom \HH(2^{m-1}) \oplus \HH(2^{m-1}) \myisom \C \otimes \HH(2^{m-1})
\end{equation}
that we exploit, with the enlarged definition of reducibility given by the second isomorphism, when the Clifford algebra reduces from $\C(2^m)$ to its subalgebra $\HH(2^{m-1})$. Let us show this with a simple example: let $V = \C^2$ and the chosen signature $(0,2)$, namely $\nu = -2 \equiv 6 \mymod{8}$. It is easy to prove \cite{Porteous_1995} that this algebra, generated by two anticommuting generators of signature $(0,2)$, is $\HH$, namely a subalgebra of $\C(2)$, and precisely the subalgebra made by matrices of the form
\opt{margin_notes}{\mynote{mbh.note: here it is not the universality property of $\myClg{}{}{V}$}}%
\begin{equation}
\label{H_matrices}
\left(\begin{array}{r r} a & -\myconjugate{b} \\ b & \myconjugate{a} \end{array}\right) \qquad a, b \in \C
\end{equation}
and the real dimension of the algebra reduces from 8 to 4 but the dimension of $S$ is unchanged and remains 4 so by the usual definition of reducibility this representation would be \emph{not} reducible. We resume the situation with:
\begin{MS_Proposition}
\label{C_subalgebras}
Given a complex space $V$ of dimension $n = 2 m$ and its Clifford algebra $\C(2^m)$ let $V$ be given the signature $\nu$ then, depending on $\prescript{}{2}\nu$, the Clifford algebra of $V$ is the subalgebra of $\C(2^m)$:
$$
\begin{array}{l l}
\R(2^m) & \mbox{\rm{for}} \quad \prescript{}{2}\nu = 1 \\
\HH(2^{m-1}) & \mbox{\rm{for}} \quad \prescript{}{2}\nu = -1 \dotinformula
\end{array}
$$
\end{MS_Proposition}
%

\noindent
So complex Clifford algebra $\C(2^m)$ \emph{always} ``reduces'' to one of its two subalgebras $\R(2^m)$ or $\HH(2^{m - 1})$ both with halved dimension with respect to $\C(2^m)$. When $\prescript{}{2}\nu = 1$ the Clifford algebra of $V$ is $\R(2^m)$ with real, Majorana, spinors. For $\prescript{}{2}\nu = -1$ the Clifford algebra is $\HH(2^{m-1})$ and so spinors are intrinsically quaternionic and $S \myisom \HH^{2^{m-1}}$. In a nutshell spinors are: either real, Majorana, spinors or quaternionic and in \emph{any case} $\C$ can leave the scene%
\footnote{a technical remark: in the real case the Clifford algebra of an even dimensional $V$ is, depending on $\prescript{}{2}\nu$, either $\R(2^m)$ or $\HH(2^{m-1})$, two non isomorphic algebras of same dimension. In the complex case, either by representation (\ref{representations}) or complexification, we obtain the same two algebras but now they are both subalgebras of $\C(2^m)$.}%
.

We remark that proposition~\ref{C_subalgebras}, that we derived from (\ref{C_R_split}) and (\ref{C_H_split}), descends also immediately from (\ref{nu_bits_def}) and the general result \cite[Proposition~13.28]{Porteous_1981}
$$
\C \otimes \myClg{}{}{\R^{k,l}} \myisom \myClg{}{}{\C^{k+l}} \myisom \C(2^m) \dotinformula
$$
In physics electromagnetism bring to use complex representations (\ref{representations}) and proposition~\ref{C_subalgebras} shows that the complex numbers, introduced ``ad hoc'', are fatally due to disappear once a signature is chosen.

For the quaternionic case we remark that with the standard isomorphism (\ref{C_to_H_isomorphism}) we can, if so we like, continue to look at spinors as at elements of $\C^{2^{m}}$, for example the familiar Dirac spinors of $\C^4$. The main change is in the matrices of $\End_\C \C^{2^m}$ that are not full fledged $\C(2^m)$ matrices but matrices representing the subalgebra $\HH(2^{m-1})$, that for example in the case of $m=1$ have the form (\ref{H_matrices}) easy to generalize to $m > 1$. Moreover even if we reduce to a quaternionic algebra, formally eliminating $\C$ algebra in (\ref{C_H_split}), there is still room for electromagnetism since in principle we can replace $i \in \C$ with any of $i, j, k \in \HH$ that square to $-1$.

We remark that quaternionic formulations of quantum mechanics are studied since long time, see \eg \cite{Giardino_2018, De_Leo_Rodrigues_1998} and references therein, in particular it has been studied and solved the problem of ``doubling of solutions'' \cite{De_Leo_1996} that can be traced to the first isomorphism of (\ref{C_H_split}).

We conclude observing that electromagnetism induces to use complex Clifford algebras but if $\prescript{}{2}\nu = 1$ complex numbers disappear completely. The other possibility is $\prescript{}{2}\nu = -1$ that, for the familiar Minkowski space with $n = 4$, leaves room for three signatures: $(4,0)$ and $(0,4)$ corresponding to $\nu \equiv 4 \mymod{8}$ and $(1,3)$ corresponding to $\nu \equiv 6 \mymod{8}$. Ruling out the two definite, Euclidean, signatures we are left with the only possibility $(1,3)$, the signature of quaternions.

This argument shows another possible application of proposition~\ref{Theta_prop}: supposing we know the dimension $n$ of $V$ and the characteristics of the involution $\Theta$ (\ref{Theta_def_2}), namely $\Theta^2$ and its parity $s$ or $\omega^2$, for example by physical motivations,
\opt{margin_notes}{\mynote{mbh.note: for example the existence of antiparticles forces to use $\rho$ involution and parity gives $\omega^2$ (more or less)}}%
than we obtain $\nu \mymod{8}$ and thus $V$ signature, a kind of indirect measurement of the physical signature of $V$.

\section{Conclusions}
\label{Conclusions}
We have shown that taking into account also the involution generated by complex conjugation the relations between $n$ and $\nu$ are the same for real and complex Clifford algebras. As a consequence once a signature is chosen for complex vector space $V$ the ``generic'' Clifford algebra $\C(2^m)$ reduces to one of its subalgebras of halved dimension $\R(2^m)$ or $\HH(2^{m-1})$ and consequently the spinors, the carriers of the representation, are either real or quaternionic. Only in this last case there is room for electromagnetism and this allows to argue in favour of signature $(1,3)$ for Minkowski space.

\newpage

\section*{Appendix: proof of proposition \ref{Theta_prop}}
\begin{proof}
\opt{margin_notes}{\mynote{mbh.ref: see log p. 703--705}}%
In the real case $\Theta = \tau$ and (\ref{general_nu_bits_derivation}) reduces to (\ref{n_bits_def}) so we need to prove just the $\C$ case and we begin proving $\prescript{}{1}\nu = \prescript{}{1}n \, s = \omega^2$; along the proof we will frequently use some signs like $s_1$ (\ref{s1_def}) and we collect here all their definitions for easy reference
\begin{equation}
\label{s_i_defs}
\begin{array}{l l l}
\omega \tau \omega^{-1} & = & s_1 \, \tau \\
\omega \rho \omega^{-1} & = & s_2 \, \rho \\
\rho \tau \rho^{-1} & = & s_3 \, \tau \\
\omega \Theta \omega^{-1} & = & s \, \Theta
\end{array}
\end{equation}
and remark that the same sign applies to the complementary relation, for example $\omega \tau \omega^{-1} = s_1 \, \tau$ implies $\tau \omega \tau^{-1} = s_1 \, \omega$.

Taking for $\omega^2$ its natural sign it is a standard exercise to get $\omega^2 = (-1)^{\frac{\nu (\nu - 1)}{2}} = \prescript{}{1}\nu$. On the other hand since $\omega = \mygenC_1 \mygenC_2 \cdots \mygenC_n$
$$
\omega^2 = (-1)^{\frac{n (n - 1)}{2}} \prod_{j = 1}^n \mygenC_j^2 = \prescript{}{1}n \, \prod_{j = 1}^n \mygen_j^2 \prod_{j = 1}^n \iota_j^2 = \prescript{}{1}n \, \prod_{j = 1}^n \mygen_j^2 (-1)^r
$$
where in the last passage we assumed that $\iota_j = i$ for $0 \le r \le n$ generators. In the last expression we recognize in $\prescript{}{1}n \, \prod_{j = 1}^n \mygen_j^2$ the value of $\omega^2$ in the real case and thus, with (\ref{n_bits_def}), $\prod_{j = 1}^n \mygen_j^2 = s_1$ and thus we have proved that in the complex case $\omega^2 = \prescript{}{1}n \, s_1 (-1)^r$. To complete the proof we need to show that $s = s_1 (-1)^r$ and by (\ref{Theta_def_2}) and (\ref{s_i_defs}) we see that $s = s_1 s_2$ since
$$
s \Theta = \omega \Theta \omega^{-1} = \omega \tau \omega^{-1} \omega \rho \omega^{-1} := s_1 s_2 \tau \rho = s_1 s_2 \Theta
$$
so to prove the thesis we must prove that $s_2 = (-1)^r$ and by its definition (\ref{s_i_defs}) $s_2 = (-1)^x$ where $x \ge 0$ is the number of generators composing $\rho$ and by (\ref{rho_explicit_def}) we see that depending on $r$ being even or odd, respectively $s_2 = (-1)^r$ and $s_2 = (-1)^{n - r} = (-1)^r$ and thus in both cases $s_2 = (-1)^r$ and this concludes the proof that in the complex case $\prescript{}{1}\nu = \prescript{}{1}n \, s = \prescript{}{1}n \, s_1 s_2 = \omega^2$.

\smallskip

We turn now to the more intricate proof that $\prescript{}{2}\nu = \prescript{}{2}n \, \Theta^2$. Also in the complex case $\tau$ is the inner element giving $\tau \mygenC_j \tau^{-1} = \mygenC_j^{\mydual}$ and this property is not affected by the redefinition $\mygenC_j = \iota_j \mygen_j$ and thus the blade $\tau_\R$ of the real case (\ref{tau_def}) is just updated in the complex case by the replacement of $\mygen_j$ by $\mygenC_j = \iota_j \mygen_j$ and thus in the complex case $\tau^2 = \tau_\R^2 (-1)^y$ where $y \ge 0$ is the number of generators entering $\tau$ for which $\iota_j = i$ but by (\ref{s_i_defs}) $(-1)^y = s_3$ and thus $\tau^2 = \tau_\R^2 s_3$. Since $\Theta = \tau \rho$ by (\ref{s_i_defs}) $\Theta^2 = s_3 \tau^2 \rho^2$ and thus $\Theta^2 = \tau_\R^2 \rho^2$ and using (\ref{n_bits_def}) we see that the relation we want to prove has taken the form
$$
\prescript{}{2}\nu = \prescript{}{2}\nu_\R \, \rho^2
$$
where by $\nu_\R$ we indicate the space signature we would get from $\mygenC_j$ setting all factors $\iota_j = 1$.

To calculate the relation between $\nu$ and $\nu_\R$ we suppose that the signature of $\nu_\R$ is $(k, l)$ and that in the complex case there are $r$ generators that are multiplied by $\iota_j = i$ of which $0 \le p \le \min{(k, r)}$ are applied to the spacelike generators $\mygen_j$ ($\mygen_j^2 = 1$). Then it is simple to see that in the complex case there will be $k + r - 2 p$ spacelike generators ($\mygenC_j^2 = 1$) and $l - r + 2 p$ timelike generators ($\mygenC_j^2 = -1$) and thus in the complex case
$$
\nu = k + r - 2 p - (l - r + 2 p) = k - l + 2 r - 4 p = \nu_\R + 2 r - 4 p \dotinformula
$$

Since we are interested in the relations between bits $\prescript{}{2}\nu$ and $\prescript{}{2}\nu_\R$ we examine the quantity $2 r - 4 p$ in binary form in the two different cases of $r$ even or odd: for $r$ even $2 r - 4 p$ can be written $r_1 0 0 - p_0 0 0$ and since $4 \equiv -4 \mymod{8}$ we obtain
\begin{equation}
\label{nu_nu_r_even}
r_1 0 0 - p_0 0 0 =
\left\{
\begin{array}{l l}
0 0 0 & \quad \mbox{iff} \quad p_0 = r_1\\
1 0 0 & \quad \mbox{iff} \quad p_0 \ne r_1
\end{array} \right. 
\end{equation}
while in the case of $r$ odd we get
\begin{equation}
\label{nu_nu_r_odd}
r_1 1 0 - p_0 0 0 =
\left\{
\begin{array}{l l}
0 1 0 & \quad \mbox{iff} \quad p_0 = r_1\\
1 1 0 & \quad \mbox{iff} \quad p_0 \ne r_1
\end{array} \right. 
\end{equation}
\opt{margin_notes}{\mynote{mbh.note: can we use these relations also for $\prescript{}{1}\nu$ ? Yes, give $\prescript{}{1}\nu = \prescript{}{1}\nu_\R \, (-1)^r$}}%
where we have examined all $4$ possible cases for $r_1, p_0$ and remembered that the relation is modulo $8$ and thus for example $2 - 4 = -2 \equiv 6 \mymod{8}$.

We proceed with the calculation of $\rho^2$, let
$$
t := \left\{
\begin{array}{l l}
r & \quad \mbox{for} \; r \; \mbox{even} \\
n - r & \quad \mbox{for} \; r \; \mbox{odd}
\end{array} \right. 
$$
and with (\ref{rho_explicit_def}) we get
\begin{equation}
\label{rho2}
\rho^2 = (-1)^{\frac{t (t - 1)}{2}} \prod_{j \in \{t\}} \mygenC_j^2 = \prescript{}{1}t \prod_{j \in \{t\}} \mygen_j^2 \, (\iota_j)^{2 t} = \prescript{}{1}t \prod_{j \in \{t\}} \mygen_j^2
\end{equation}
where the last passage is justified by the fact that for $r$ even $(\iota_j)^{2 t} = (-1)^r = 1$ and also for $r$ odd $(\iota_j)^{2 t} = (1)^{2 (n - r)} = 1$.

By $t$ definition for $r$ even $\prescript{}{1}t = \prescript{}{1}r$ while for $r$ odd $t = n - r$ that in binary form is $n_2 n_1 0 - r_2 r_1 1$ and $-r$ in binary form is $\overline{r}_2 \overline{r}_1 1$ (where $\overline{r}_i$ is the complementary bit of $r_i$) and thus in binary form $t = n + (-r)$ reads
$$
t_2 t_1 t_0 = n_2 n_1 0 + \overline{r}_2 \overline{r}_1 1
$$
so that, since there are no carries from the least significant bit, $t_1 = n_1 + \overline{r}_1$ and we can summarize these results with
$$
\prescript{}{1}t := \left\{
\begin{array}{l l}
\prescript{}{1}r & \quad \mbox{for} \; r \; \mbox{even} \\
- \prescript{}{1}r \prescript{}{1}n & \quad \mbox{for} \; r \; \mbox{odd.}
\end{array} \right. 
$$

It remains to calculate the factor $\prod_{j \in \{t\}} \mygen_j^2$ that depends on the number of generators in $\rho$ that are antisymmetric but by (\ref{s_i_defs}) $\tau \rho \tau^{-1} = s_3 \rho$ and this involution replaces the generators with their dual so that we conclude that $s_3 = \prod_{j \in \{t\}} \mygen_j^2$.

With the definitions of $\rho$ (\ref{rho_explicit_def}), $r$ and $p$ we see that for $r$ even there are $r - p$ antisymmetric generators in $\rho$ and thus $s_3 = (-1)^{r - p} = (-1)^{p} = \prescript{}{0}p$ while for $r$ odd $\rho$ has $(n - r) - (k - p)$ antisymmetric generators and thus
$$
s_3 = (-1)^{r} (-1)^{k} (-1)^{p} = - \prescript{}{0}k \prescript{}{0}p = - s_1 \prescript{}{0}p
$$
since by (\ref{tau_def}) and (\ref{s_i_defs}) $\prescript{}{0}k = \prescript{}{0}l = s_1$. In summary
$$
s_3 = \left\{
\begin{array}{l l}
\prescript{}{0}p & \quad \mbox{for} \; r \; \mbox{even} \\
- s_1 \prescript{}{0}p & \quad \mbox{for} \; r \; \mbox{odd}
\end{array} \right. 
$$
and putting these results together we get
\begin{equation}
\label{rho2_def}
\rho^2 = \prescript{}{1}t \, s_3= \left\{
\begin{array}{l l}
\prescript{}{1}r \prescript{}{0}p & \quad \mbox{for} \; r \; \mbox{even} \\
\prescript{}{1}r \prescript{}{0}p \prescript{}{1}n \, s_1 = \prescript{}{1}r \prescript{}{0}p \prescript{}{1}\nu_\R & \quad \mbox{for} \; r \; \mbox{odd}
\end{array} \right. 
\end{equation}
where to obtain the second relation for $r$ odd we used the real case relation (\ref{n_bits_def}) for $\prescript{}{1}\nu_\R$.

We can now verify the relation under scrutiny, $\prescript{}{2}\nu = \prescript{}{2}\nu_\R \, \rho^2$, separately in the two cases of $r$ even or odd; for $r$ even by (\ref{rho2_def}) we get that $\prescript{}{2}\nu = -\prescript{}{2}\nu_\R$ if and only if $p_0 \ne r_1$ that, compared with (\ref{nu_nu_r_even}), confirms that in this case $\prescript{}{2}\nu$ and $\prescript{}{2}\nu_\R$ are actually opposite.

For $r$ odd we distinguish two cases: for $\prescript{}{0}p = \prescript{}{1}r$ by (\ref{rho2_def}) $\rho^2 = \prescript{}{1}\nu_\R$ and this coincides with the result given by (\ref{nu_nu_r_odd}): $\prescript{}{2}\nu \ne \prescript{}{2}\nu_\R$ if and only if a carry is generated in the second bit with $\prescript{}{1}\nu_\R$. For $\prescript{}{0}p \ne \prescript{}{1}r$ then $\rho^2 = - \prescript{}{1}\nu_\R$ and thus $\prescript{}{2}\nu$ and $\prescript{}{2}\nu_\R$ are opposite if and only if $\prescript{}{1}\nu_\R = 1$ and this is again confirmed by (\ref{nu_nu_r_odd}) since $\prescript{}{2}\nu \ne \prescript{}{2}\nu_\R$ if and only if a carry is \emph{not} generated in the second bit and thus iff $\prescript{}{1}\nu_\R = 1$. This completes the proof that $\prescript{}{2}\nu = \prescript{}{2}\nu_\R \, \rho^2$.
\end{proof}


\opt{x,std,AACA}{
\bibliographystyle{plain} 

\bibliography{mbh}

\begin{thebibliography}{10}

\bibitem{Budinich_2016}
Marco Budinich.
\newblock On spinors transformations.
\newblock {\em Journal of Mathematical Physics}, 57(7):071703--1--11, July
  2016.
\newblock arXiv:1603.02181 [math-ph] 7 Mar 2016.

\bibitem{Budinich_2016b}
Marco Budinich.
\newblock {On Clifford Algebras and Binary Integers}.
\newblock {\em Advances in Applied Clifford Algebras}, 27(2):1007--1017, 2017.
\newblock arXiv:1605.07062 [math-ph] 23 May 2016.

\bibitem{BudinichP_1988e}
Paolo Budinich and Andrzej~Mariusz Trautman.
\newblock {\em {The Spinorial Chessboard}}.
\newblock Trieste Notes in Physics. Springer-Verlag, Berlin Heidelberg, 1988.

\bibitem{Cartan_1913}
{\'{E}}lie Cartan.
\newblock Les groupes projectifs qui ne laissent invariante aucune
  multiplicit{\'e} plane.
\newblock {\em Bulletin de la Soci{\'e}t{\'e} Math{\'e}matique de France},
  41:53--96, 1913.

\bibitem{Cartan_1937}
{\'{E}}lie Cartan.
\newblock {\em The Theory of Spinors}.
\newblock Hermann, Paris, 1966.
\newblock first edition: 1938 in French.

\bibitem{Chevalley_1954}
Claude~C. Chevalley.
\newblock {\em Algebraic Theory of Spinors}.
\newblock Columbia University Press, New York, 1954.

\bibitem{Conrad_2008}
Keith Conrad.
\newblock Complexification, 2008.
\newblock http://www.math.uconn.edu/ \~{
  }kconrad/blurbs/linmultialg/complexification.pdf (May 2018).

\bibitem{Dabrowski_1988}
Ludwik Dabrowski.
\newblock {\em {Group Actions on Spinors}}.
\newblock Lecture Notes. Bibliopolis, Napoli, 1988.

\bibitem{De_Leo_1996}
Stefano {De Leo}.
\newblock A one-component dirac equation.
\newblock {\em International Journal of Modern Physics A}, 11(21):3973--3985,
  1996.

\bibitem{De_Leo_Rodrigues_1998}
Stefano {De Leo} and Waldyr~Alves {Rodrigues Jr.}
\newblock Quaternionic electron theory: Dirac's equation.
\newblock {\em International Journal of Theoretical Physics}, 37(5):1511--1529,
  May 1998.

\bibitem{Giardino_2018}
Sergio Giardino.
\newblock {Quaternionic quantum mechanics in real Hilbert space}, March 2018.
\newblock arXiv:1803.11523 [quant-ph] 30 Mar 2018.

\bibitem{Granville_1997}
Andrew Granville.
\newblock {Arithmetic properties of binomial coefficients. I. Binomial
  coefficients modulo prime powers.}
\newblock In {\em Canadian Mathematical Society Conference Proceedings},
  volume~20, pages 253--276, Providence, RI, 1997. American Mathematical
  Society.

\bibitem{Porteous_1981}
Ian~Robertson Porteous.
\newblock {\em Topological geometry}.
\newblock Cambridge University Press, {II} edition, 1981.

\bibitem{Porteous_1995}
Ian~Robertson Porteous.
\newblock {\em Clifford Algebras and the Classical Groups}.
\newblock Cambridge Studies in Advanced Mathematics: 50. Cambridge University
  Press, 1995.

\bibitem{Trautman_1998}
Andrzej~Mariusz Trautman.
\newblock On complex structures in physics.
\newblock In Alex Harvey, editor, {\em {On Einstein's path: essays in honor of
  Engelbert Schucking}}, chapter~34, pages 487--495. Springer Verlag, New York,
  1999.

\bibitem{Varlamov_2001}
Vadim~Valentinovich Varlamov.
\newblock Discrete symmetries and clifford algebras.
\newblock {\em International Journal of Theoretical Physics}, 40(4):769--805,
  April 2001.

\end{thebibliography}
}

\opt{arXiv,JMP}{
%
%

%
%
}

\opt{final_notes}{
\newpage

\section*{Things to do, notes, etc.......}

\noindent General things to do:
\begin{itemize}
\item ...
\end{itemize}

\section{Unused parts}
\label{Unused_parts}

\begin{MS_Proposition}
\label{CbarC_prop}
$$
\myconjugate{C} C = \rho^2
$$
\end{MS_Proposition}
\begin{proof}
Old, complicate, proof goes: $\myconjugate{C} C = \myconjugate{\gamma(\rho)} \gamma(\rho) = \myconjugate{\gamma}(\rho) \gamma(\rho)$ by the standard definition of $\myconjugate{\gamma}$ representation. Then $\myconjugate{C} C = \gamma({\myconjugate{\rho}}) \gamma(\rho)$ since $\rho$ is given by generators (\ref{rho_explicit_def}) and for them $\myconjugate{\gamma}(\mygenC_j) = \gamma(\myconjugate{\mygenC}_j)$. Thus $\myconjugate{C} C = \gamma(\myconjugate{\rho} \rho) = \gamma(\rho^2) = \rho^2$ since for $n$ even, $\myconjugate{\rho} \rho = \rho^2$: with conventions used in (\ref{rho2}) and $\myconjugate{\iota} \iota = 1$ we get
$$
\myconjugate{\rho} \rho = \prescript{}{1}t \prod_{j \in \{t\}} \myconjugate{\mygenC}_j \mygenC_j = \prescript{}{1}t \prod_{j \in \{t\}} \mygen_j^2 \, (\myconjugate{\iota} \iota)^t = \prescript{}{1}t \prod_{j \in \{t\}} \mygen_j^2
$$
that coincides with $\rho^2$ (\ref{rho2}) and proves the thesis.
\end{proof}

\begin{MS_Proposition}
\label{C_mat_prop}
$$
\C(2^m) \myisom \HH(2^{m-1}) \oplus \HH(2^{m-1})
$$
\end{MS_Proposition}
\begin{proof}
We proceed by induction on $m$, for $m = 1$ the generic element of $\C(2)$ has the form $\left(\begin{array}{r r} a & b \\ c & d \end{array}\right) \myisom \left(\begin{array}{r r} a & 0 \\ c & 0 \end{array}\right) \oplus \left(\begin{array}{r r} 0 & b \\ 0 & d \end{array}\right)$ each of the term being isomorphic to $\C^2$ and thus to $\HH$. Let the proposition be true for $m - 1$ then the generic element of $\C(2^m)$ has the form $\left(\begin{array}{r r} a & b \\ c & d \end{array}\right)$ where $a,b,c,d \in \C(2^{m-1})$ and thus by hypothesis each of them is isomorphic to $\HH(2^{m-2}) \oplus \HH(2^{m-2})$. Thus $\C(2^m)$ is isomorphic to the direct sum of eight $\HH(2^{m-2})$ that can be collected in two groups of four proving the thesis.
\end{proof}

\section{old section: Deriving the space signature of a Clifford algebra}
\label{Space_signature_real}
We exploit these results to derive the vector space signature from the characteristics of the fundamental involutions of the algebra.

As a warming-up exercise let us suppose that we have been given a real Clifford algebra of a vector space $V$ and we know its dimension $n$; we thus know the binary expansion of $n$ namely the bits $\prescript{}{2}n, \prescript{}{1}n$ and $\prescript{}{0}n$ and we know also $\prescript{}{0}\nu = \prescript{}{0}n$. At this point if we are given the values of $\tau^2$ and either $\omega^2$ or $s_1$ by (\ref{n_bits_def}) we can calculate $\nu$, that together with $n$, gives the exact signature of $V$.

To proceed we review briefly some fundamental results and their role in Clifford algebras and in the problem at hand:
\begin{itemize}
\item any matrix algebra of size $2^n$ contains $2 n + 1$ anticommuting matrices with square equal to $\pm \Identity$ of which $n$ are antisymmetric and $n+1$ are symmetric and at most one of the symmetric matrices is diagonal. Moreover there cannot be more than $2 n + 1$ of these matrices otherwise contradictions would arise in Clifford algebra (find precise ref. in \cite{Porteous_1995}).%
\opt{margin_notes}{\mynote{mbh.ref: log p. 671}}%
\item by Wedderburn-Artin theorem all Clifford algebras are matrix algebras of size $2^r$, for some $r$, over one of the $5$ division algebras of figure~\ref{nu_drawing} and in particular $\R$ or $\HH$ for even dimensional spaces and $\R \oplus \R$, $\HH \oplus \HH$ or $\C$ for odd dimensional ones;
\item the Clifford algebra of any quadratic space $V$, by injections $\F \to \myClg{}{}{V}: x \to x \Identity$ and $V \to \myClg{}{}{V}$, the so called Clifford map, contains ``copies'' of the field $\F$ and of the vector space $V$. Whereas the field injection is unique, there are various possibilities for the injections of the vector space $V$. In any case the Clifford map always selects a subset of the anticommuting matrices of ``its'' matrix algebra and it is customary to identify $V$ with its image in $\myClg{}{}{V}$ and an orthonormal base of $V$ with the anticommuting matrices selected by the chosen injection. We remark that these matrices are not ``representations'' but members of the algebra itself;
\item in any real Clifford algebra the anticommutator $\anticomm{\mygen_{i}^{\mydual}}{\mygen_j} = 2 \delta_{i j}$ \cite[Proposition~13.27]{Porteous_1981} and thus it follows that
$$
\tau \mygen_{i} \tau^{-1} = \mygen_{i}^{\mydual} = \mygen_{i}^{-1} = \mygen_{i}^{3} \dotinformula
$$
\end{itemize}

\begin{MS_Proposition}
\label{blades_prop}
In any Clifford algebra given any homogeneous product of generators $\rho = \mygen_{i} \mygen_{j} \cdots \mygen_{k}$ (sometimes called a ``blade'')
\opt{margin_notes}{\mynote{mbh.ref: log p. 646}}%
\opt{margin_notes}{\mynote{mbh.probs: tricky notation: here $ \rho^{\mydual}$ represents `reversed' order and not the $\tau$ automorphism}}%
$$
\rho^{-1} = \rho^{\mydual} = \rho^2 \rho = \pm \rho \dotinformula
$$
\end{MS_Proposition}
\begin{proof}
For any $\rho = \mygen_{i} \mygen_{j} \cdots \mygen_{k}$ we have $\rho^{\mydual} = \mygen_{k}^{\mydual} \cdots \mygen_{j}^{\mydual} \mygen_{i}^{\mydual}$ and since $\mygen_{i}^{\mydual} \mygen_{i} = \Identity$ it follows $\rho^{\mydual} \rho = \rho \rho^{\mydual} = \Identity$; moreover since $\rho^2 = \pm \Identity$ and thus $\rho^2 \rho = \rho^{-1}$, by the unicity of the inverse, the result is proved.
\end{proof}

With this results applied to $\omega$ and $\tau$ it follows that knowing \eg $\omega^2$ is equivalent to knowing $\omega^{\mydual}$ but whereas the first depends on the signatures of the generators $\mygen_{i}^{2} = \pm 1$ the second depends only on the symmetry of $\omega$ and is more amenable to be used in cases in which the signature is unknown, \eg in complex linear spaces. In this fashion we will be able to determine space signatures from the knowledge of $\omega^{\mydual}$ and $\tau^{\mydual}$ that \emph{do not} depend on generator signatures but only on symmetry of the main automorphisms (or involutions ?) of the algebra.

We tackle now the simpler case of a Clifford algebra over the complex field $\C$: here there exists only one possible ``variable'', $n$ and it is sufficient to know if $n$ is even or not, namely $n \mymod{2} = n_0$, to know if the algebra is respectively central simple or semisimple. Still knowing $n$ we know also the values of the two other bits $n_1$ and $n_2$ and we know that they are related by (\ref{general_nu_bits_derivation}) to $\nu$ even if it is not yet clear what $\nu$ could represent in this case.

Nevertheless for Clifford algebras of a complex vector space the fundamental automorphisms are defined and, if $n$ is even, they are inner and thus the elements $\omega$ and $\tau$ are properly defined and by studying their symmetry, namely, $\omega^{\mydual}$ and $\tau^{\mydual}$ we can exploit the just seen properties to obtain $\nu$. A similar result is obtained by Varlamov \cite{Varlamov_2001} that essentially by the automorphism group of the Clifford algebra derives the sign $s$ and relates it to $\prescript{}{1}n$ from which, by (\ref{general_nu_bits_derivation}), $\prescript{}{1}\nu = s \, \prescript{}{1}n$.
\opt{margin_notes}{\mynote{mbh.ref: see leaf S6; see \cite{Varlamov_2001} (P\_667) pages 8 and 31--32}}%

The same procedure can be done when we complexify a real Clifford algebra since it is well known that, for even dimensional spaces $k + l = n = 2 m$ \cite[Proposition~13.28]{Porteous_1981}
$$
\myClf{k}{l}{\R} \otimes \C \myisom \myClC{2 m}{}{} \myisom \C(2^m)
$$
and so any information of the signature of the real algebra is lost by complexification since all real Clifford algebras merge in only one complex Clifford algebra as is customary to do \eg in Dirac algebra.

Is there some information on the initial real signature in complex Dirac algebra ?.

First of all we reproduce the main argument used to support the statement that when $V = \C^n$ there are no signatures: the reason is that by (\ref{f_j_def}) any generator can be freely given any sign.

But in real space the signature is in one to one correspondance with the symmetry of the generator under the dual symmetry since in the real case we have
$$
\mygen_{i}^{\mydual} = \mygen_{i}^{-1} = \mygen_{i}^2 \mygen_{i} = \pm \mygen_{i}
$$
and the symmetry under the dual operation is maintained also in complexified algebra. As en example consider the generator $\mygen_{2} = \epsilon$ of the $\myClf{1}{1}{\R} \myisom \R(2)$ that represents the timelike coordinate and is antisymmetric since $\mygen_{2}^{\mydual} = - \mygen_{2}$; when complexifying to $\C(2)$ then matrix $\epsilon$ is still antisymmetric but it can be done spacelike considering $i \epsilon$ as is customary to do for Pauli matrices.

If we assume that the more fundamental property of the generators is their symmetry properties these remain unaffected when complexifying.

If $n$ is even the algebra is always central simple and there must be inner elements for the main involutions.

It follows that also in complex Clifford algebra and in complexified Clifford algebra we can define as in the real case the inner elements $\omega, \tau$ and $\rho$ giving the three fundamental involutions (commented citation Varlamov\_2015)
\begin{eqnarray*}
\omega \mygen_i \omega^{-1} & = & - \mygen_i \qquad \qquad \forall \: 1 \le i \le n \\
\tau \mygen_{i} \tau^{-1} & = & \mygen_{i}^{\mydual} \\
\rho \mygen_{i} \rho^{-1} & = & \myconjugate{\mygen_{i}} 
\end{eqnarray*}
and assuming that we know the three least significant bits of $n$ in the complex or complexified case we can give them this meaning: $\prescript{}{0}n$ as usual conveys the information wether the dimension of the space $V$ is even or odd and consequently wether the Clifford algebra is central simple or not. As already pointed out in this case $\prescript{}{1}n = \omega^2$ if $\mygen_{i}^2 = 1$ for all $i$.

We can thus use (\ref{general_nu_bits_derivation}) to calculate the bits of $\nu$ of an hypothetical real Clifford algebra with same properties: it easily follows that $\prescript{}{2}\nu = 1$, $\prescript{}{1}\nu = \prescript{}{1}n = \omega^2$ and $\prescript{}{0}\nu = \prescript{}{0}n = 1$. This means that this associated real Clifford algebra has necessarily $\nu \in \{ 0, 2 \}$ and is thus isomorphic to $\R(2^m) \myisom \myClf{m}{m}{\R} \myisom \myClf{m+1}{m-1}{\R}$ (is this the real ``part'' of the complex Clifford algebra ??? another ``proof'': $\C(4)$ has real part $\R(4)$ and thus $\nu \in \{ 0, 2 \}$).

To do this we first observe that in the complexification of $\myClf{k}{l}{\R}$ one chooses to go to the Clifford algebra of vector space $\myconjugate{\C}^{n,0}$ since we need to have positive definite inner product on spinor space $S$ if we want to have positive definite probabilities.

The second observation is that choosing the map $C: S \to \myconjugate{S}$ we are actually defining a real structure on $S$ and this gives some informations of the submerged signature.

} 

\end{document}